# A Circuit-Theoretic Approach to State Estimation

Shimiao Li, Amritanshu Pandey, Soummya Kar, Larry Pileggi

Dept. of Electrical and Computer Engineering
Carnegie Mellon University
Pittsburgh, PA

*Abstract*— Traditional state estimation (SE) methods that are based on nonlinear minimization of the sum of localized measurement error functionals are known to suffer from non-convergence and large residual errors. In this paper we propose an equivalent circuit formulation (ECF)-based SE approach that inherently considers the complete network topology and associated physical constraints. We analyze the mathematical differences between the two approaches and show that our approach produces a *linear* state-estimator that is mathematically a quadratic programming (QP) problem with closed-form solution. Furthermore, this formulation imposes additional topology-based constraints that provably shrink the feasible region and promote convergence to a more physically meaningful solution. From a probabilistic viewpoint, we show that our method applies prior knowledge into the estimate, thus converging to a more physics-based estimate than the traditional observation-driven maximum likelihood estimator (MLE). Importantly, incorporation of the entire system topology and underlying physics, while being linear, makes ECF-based SE advantageous for large-scale systems.

## I. INTRODUCTION

State estimation (SE) is a central part of power system operation and security. To infer the operating point of the grid, the state of the system is estimated based on various measurements within the supervisory control and data acquisition (SCADA) system that is installed within the network. The output solution of this state estimator is critical, since it is fed into the real-time contingency analysis (RTCA) module, as well as the real-time dispatch module. A non-functioning SE will result in failure of these processes.

Traditionally, the state variables for the SE have been voltage phasor magnitudes and angles at all buses. To estimate them correctly, the classical approach proposed by [1] formulates SE as a minimization of weighted least-squares (WLS) error of measurement noise. However, the intrinsic nature of WLS problem, the unavoidable noise corruption, as well as the violation of the assumed noise distribution property, collectively create problems with this traditional method.

As depicted in Figure 1, the major limitation of the traditional SE method lies in the following aspects:
- **Non-convergence and high residual (convergence to local minima or saddle point) problem**. The WLS objective expresses each measurement by a function of state variables and minimizes the mismatch. However, due to the nonlinear relationships between measurements and states, the WLS-based SE is a highly nonlinear nonconvex problem solvable by iterative methods. During the updates, each iteration solves a 'new' system that is a linearization of the target system at the current solution, and therefore, it lacks real physical meaning. Once there is a bad initial guess, iterative methods can deviate largely from reality, leading to non-convergence or convergence to local minima or saddle points.

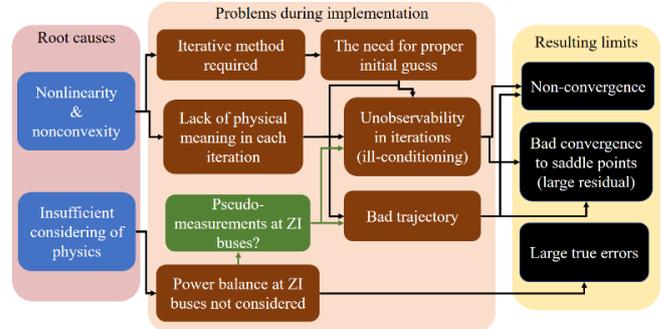

Figure 1. Limitations of WLS-based SE

- **Insufficient real-world physics in SE**. Ideally, SE would provide grid state estimates with minimal deviation from the true grid state, which should result in feasible power flow constraints. However, in a WLS formulation, such complete network power balance is not fully considered, and the desired outcome is approximated by a minimization of residual between estimated states and measurement. In this case, when some measurements are corrupted by noise, missing completely, or grossly inaccurate, the resulting solution is still forced to match the measurements to the largest extent, but thereby fails to sufficiently minimize the true deviation; i.e., the true error between system states and obtained estimates. We note that, depending on model complexity and noise statistics, other types of M-estimators [2] may be used to resolve this issue, which is a topic of general interest. We will partially address this issue in this paper. We show that the topological constraints of our formulation effectively correspond to adding a regularization penalty in the residual minimization objective, thus promoting solutions that adhere to physical network constraints (Section III-C).

Earlier research has tried to address the aforementioned limitations. Firstly, SEs with just PMUs have been developed that are in linear in complexity [3]. However, the penetration of PMUs compared to RTUs remains low. To deal with the phasor angle measurements and consider SE with both conventional RTU meters and PMUs, some novel methods rely on a transformed use of angle meters [4], a hybrid method consisting of two separate estimators [5], and some linearly remodeled PMUs [6]. However, in estimators for combined measurements, the nonlinear nature of conventional measurements and the resulting convergence problems still exist. Similarly, to improve upon the true residual problem, more grid physics has been incorporated into the problem through modeling of zero injection nodes through pseudo measurements [7]. However, adding these additional terms with extremely high weights has resulted in matrix ill-conditioning and corresponding non-convergence. The constrained Hachtel optimization approach [8] is also proposed to solve the SE problem with zero-injection (ZI) nodes, but it has been found hard to solve due to the non-convexity of the optimization problem.

To tackle all the limitations of the WLS-based method, we proposed an equivalent circuit formulation (ECF) based approach to state-estimation [9][11][15]. This paper builds on that and presents a comprehensive comparison between the traditional WLS method and the ECF-based SE approach. For the latter approach, we introduce circuit-based models for the phasor measurement units (PMUs) as well as novel line flow meters from any RTU or PMU devices. The main contribution of this paper is to mathematically demonstrate how our approach addresses the root causes of the aforementioned limitations:

- Resolve all convergence-related issues by formulating SE as a Quadratic Programming (QP) problem with closed-form solution, thereby reaching a global minimum.
- Enforce a more physically meaningful solution by considering the complete network topology.

Section 2 begins with a background overview of ECF formulation. Section 3 shows the mathematical comparison in terms of measurement models, problem formulation and probabilistic illustration. Section 4 presents some experimental results. Finally, section 5 documents the conclusions.

## II. BACKGROUND

### A. Notation

Table 1 shows the symbols used in this paper.

TABLE 1: SYMBOLS AND DEFINITIONS

| Symbol | Interpretation |
|---|---|
| $V^R, V^I, I^R, I^I$ | Real/imaginary voltage/current; |
| $|V|, \theta, P, Q$ | Voltage magnitude/angle; active/reactive power |
| $x$ | State variable vector, $x = [V^R, V^I]$ |
| $z$ | Measurements |
| $n, n^R, n^I$ | Noise term; real and imaginary noise term |
| $z_V^R, z_V^I, z_I^R, z_I^I$ | Measurement of Real/imaginary voltage/current |
| $z_{|V|}, z_P, z_Q$ | Measurement of $|V|, P, Q$ |
| $I(x) = 0$ | Network balance (KCL) equations |
| $Y$ | Admittance Matrix s.t. I=YV |
| $Y_i^R, Y_i^I$ | Admittance vector at bus i: $I_i^R = Y_i^R x, I_i^I = Y_i^I x$ |
| $G, B$ | Conductance, susceptance |

### B. Equivalent Circuit Formulation (ECF)

A circuit-theoretic formulation for power flow and grid optimizations was developed in [12]-[13]. Instead of describing components with 'PQV' parameters, this framework models each component within the power grid as an electrical circuit element characterized by its I-V relationship. For computational analyticity, the complex relationships are split into real and imaginary sub-circuits whose nodes corresponds to power system buses [12]-[13]. Table 2 shows a simple comparison between the traditional PQV formulation and ECF.

TABLE 2: COMPARISON BETWEEN FORMULATIONS

| Property | Comparison | |
|---|---|---|
| | PQV formulation | ECF (I-V formulation) |
| Coordinate | Polar | Rectangle |
| State variable | $|V|, \theta$ | $V^R, V^I$ |
| Network balance | Zero power mismatch | Zero current mismatch |
| Governing equations | Power balance at ∀bus | KCL equations at ∀bus |

Under this ECF framework, the network equations are intrinsically linear due to I-V relationships through Ohm's law and the non-linearities exist due to load and generation models. In case of SE these generation and load models are replaced via measurement models and therefore with linear models for injections and line-flow measurement devices, this framework results in a QP problem with closed form solution. The approach and models are further discussed below.

## III. METHODS AND COMPARISONS

### A. Models: Linear Models to improve convergence

Compared to the traditional SE, which nonlinearly represents the relationship between each measurement and state variables, the ECF-based develops linear models for measurements and replaces them in the aggregated circuit via *Substitution Theorem*. These models are flexible and can be added, removed or replaced to represent any changes within the grid.

In the following subsection we derive a new linear PMU model to overcome some of the challenges of the existing PMU model in ECF formulation [9]. We also briefly discuss the linear RTU model from [11] and demonstrate their advantages in improving convergence.

#### 1) Phasor Measurement Units (PMU):

As a synchronized measurement device, a PMU can provide real-time meter readings of voltage and injection current phasors in rectangular coordinates: $z_V^R, z_V^I, z_I^R, z_I^I$.

By substitution theorem, we can safely develop a new PMU model (shown in Figure 2). The model is represented by independent current sources taking value of the measured real and imaginary current. To further consider measurement errors such that KCL is always satisfied, we attach additional slack current sources $n^R, n^I$ to represent the noise term such that:

$$I^R = z_I^R + n^R \quad (1)$$

$$I^I = z_I^R + n^I \quad (2)$$

Meanwhile, the observation of voltage phasor indicates the state variables $V^R, V^I$ should be close to $z_V^R, z_V^I$, with the mismatch representing measurement errors. Consequently, we estimate state variables by minimizing the noise term and the voltage mismatch:

$$\min_{V^R, V^I, n^R, n^I} w(||n^R||_2^2 + ||n^I||_2^2) + w(||V^R - z_V^R||_2^2 + ||V^I - z_V^I||_2^2) \quad (3)$$

*s.t. Current balance (KCL equations) at buses*

The problem description above results in a provably convergent QP formulation for PMU models.

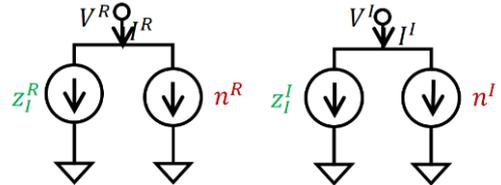

Figure 2. PMU model.

To effectively represent the difference in accuracy of individual PMU devices in the grid through weights ($w$), voltages and currents are captured separately in this model (see (3)), to address challenges of the PMU model [9].

*2) Remote terminal unit (RTU):*

RTU provides measurements of voltage magnitude $z_{|V|}$, power injection $z_P$ and $z_Q$ at a measured bus. In ECF, despite no direct phasor measurements, these observations can be mapped into linear circuit formulation [11][15] using the following relationship between bus voltages and injection currents:

$$I^R = \frac{P}{|V|^2}V^R + \frac{Q}{|V|^2}V^I \quad (4)$$

$$I^I = \frac{P}{|V|^2}V^I - \frac{Q}{|V|^2}V^R \quad (5)$$

and a linear RTU model can be developed accordingly, with resister and controlled current source parameters $G$ and $B$. Similar to the PMU model, we add an additional current source $n^R, n^I$ to capture the mismatch and represent the bus errors:

$$I^R = GV^R + BV^I + n^R \quad (6)$$
$$I^I = GV^I - BV^R + n^I \quad (7)$$

where $G = \frac{z_P}{z_{|V|}^2}, B = \frac{z_Q}{z_{|V|}^2}$

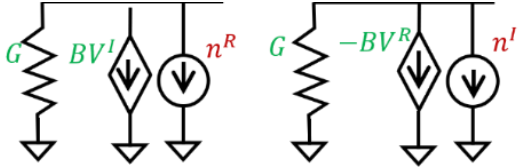

Figure 3. RTU model.

With this model, we estimate the state variables by minimizing the noise term:

$$\min_{V^R, V^I, n^R, n^I} \|n^R\|_2^2 + \|n^I\|_2^2 \quad (8)$$

s.t. *Current balance (KCL equations) at buses*

This creates a linear model for the measured RTU information.

*3) Line flow measurements of RTU and PMU*

RTU devices can also measure flow on one or more lines adjacent to the bus. This paper establishes their linear models in a way consistent with injection models. To model line flow measurements, we use an additional control circuit having injections at the measured value. The control circuits are coupled with the main circuit by voltage control. The models are shown in Figure 4, where $G = \frac{z_{Pline}}{z_{|V|}^2}, B = \frac{z_{Qline}}{z_{|V|}^2}$ and $I_{line}^{R/I}$ can be linearly expressed by ECF branch models. PMU flow model is previously discussed in [15].

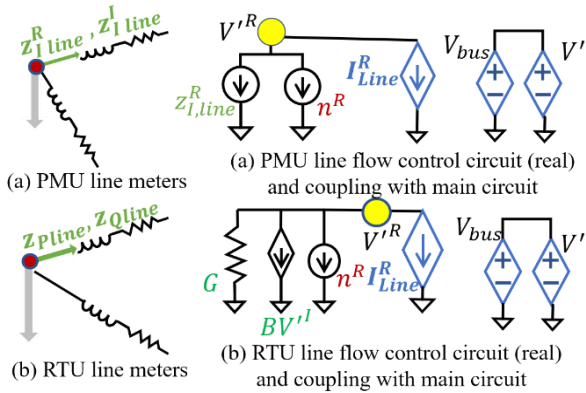

Figure 4. Linear model for line flow meters

## B. Problem formulation: a convex problem with topology-based constraints to shrink feasible set

Next, we present mathematical formulations of the ECF-based and traditional SE problems in order to compare them.

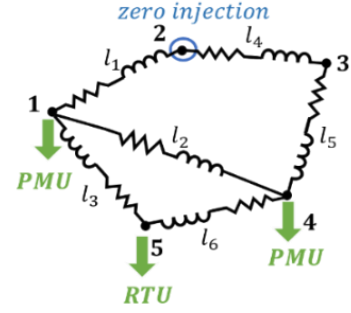

Figure 5. 5-bus case example.

Consider the simple IEEE 5-bus case shown in Figure 5. Bus #2 is a ZI bus. Suppose we obtain PMU measurements on bus 1, 4 and RTU measurements on bus 5.

*1) Traditional WLS-based method*

For a traditional SE approach, measurements are modeled by:

$$z_i = f_i(x) + n_i \quad (9)$$

To estimate state variables $x$, the traditional approach is to minimize a weighted least square of the measurement errors:

$$\min_X \sum_i w_i (z_i - f_i(x))^2 \quad (10)$$

where $f_i(x)$ is determined by Kirchoff's law and line admittances in terms of power mismatch equations, and $f_i$ is nonlinear for some *i*. This minimization has no closed form solution and is solved iteratively. With the use of polar coordinates for this traditional SE formulation, the measured phasor angle is generally expressed with the $\tan^{-1}()$ function, thus adding to any convergence difficulty.

*2) ECF-based method*

In ECF, with the linear component models proposed in Section III.A and in [15] we optimize to find the $x$ that results in minimized error while satisfying the power system balance:

$$\min_{X, n, I_{NA}} \sum_{i \in \{1,4,5\}} w_i[(n_i^R)^2 + (n_i^I)^2]$$

$$+ \sum_{i \in \{1,4\}} w_i[(V_i^R - z_{Vi}^R)^2 + (V_i^I - z_{Vi}^I)^2]$$

Subject to:
PMU buses #$i \in \{1,4\}$
$$Y_i^R x + z_{I,i}^R + n_i^R = 0 \quad (11)$$
$$Y_i^I x + z_{I,i}^I + n_i^I = 0 \quad (12)$$

RTU bus #5:
$$Y_5^R x + G_5 V_5^R + B_5 V_5^I + n_5^R = 0 \quad (13)$$
$$Y_5^I x + G_5 V_5^I - B_5 V_5^R + n_5^I = 0 \quad (14)$$

Zero-injection (ZI) bus #2:
$$Y_2^R x = 0 \quad (15)$$
$$Y_2^I x = 0 \quad (16)$$

Not measured bus #3:
$$Y_5^R x + I_{NA}^R = 0 \quad (17)$$
$$Y_5^I x + I_{NA}^I = 0 \quad (18)$$

At each measurement bus, the error terms $n^R, n^I$ can be expressed linearly with $x$. Looking at the non-measured bus (bus #3), its slack variable $I_{NA}^R, I_{NA}^I$ is purely determined by the value of state variables to make equations (17)(18) hold, thereby making no contribution to constraining and estimating $x$. Thus, we can safely remove constraints of non-measured buses and replace $n^R, n^I$ with its linear function of $x$. Then the problem is translated to the following equivalent form:

$$\min_{X, n, I_{NA}} \sum_{i \in \{1,4\}} w_i \left( \left\| Y_i^R x + z_{I,i}^R \right\|_2^2 + \left\| Y_i^I x + z_{I,i}^I \right\|_2^2 \right)$$
$$+ \sum_{i \in \{1,4\}} w_i \left( \left\| V_i^R - z_{Vi}^R \right\|_2^2 + \left\| V_i^I - z_{Vi}^I \right\|_2^2 \right)$$
$$+ w_5 \left\| Y_5^R x + G_5 V_5^R + B_5 V_5^I \right\|_2^2$$
$$+ w_5 \left\| Y_5^I x + G_5 V_5^I - B_5 V_5^R \right\|_2^2$$

Subject to zero-injection bus #2:

$$Y_2^R x = 0 \quad (19)$$
$$Y_2^I x = 0 \quad (20)$$

This formulation provides a clear layout for comparison between the traditional and ECF formulation. While the traditional SE minimizes a weighted least square error over the entire vector space, the ECF-based method is mathematically equivalent to a traditional WLS method plus some additional constraints. These constraints are independent of measurement data and are accurate system topology information that reflects the network balance on the ZI buses and any unmeasured buses with forecasted pseudo-measurements. Mathematically, the additional constraints imposed on the WLS problem shrinks the feasible space to a smaller physical region that contains an optimal solution satisfying network balance (at ZI buses). As the system grows larger with more ZI buses and other pseudo-measurements, the ECF-based method imposes an increased number of constraints to effectively localize the feasible space to a physical region. Consequently, the solution ends up being more physically meaningful.

Most importantly, owing to the linear models for each measurement device, the objective function above can be safely converted to a quadratic form $x^T Q x + Bx + C$ with $Q$ positive semi-definite. Along with linear constraints, the ECF-based SE is provably a convex problem, or more specifically, **a quadratic programming (QP)** that always converges to a global optimum of the loss function. In contrast, the traditional SE formulation results in a nonlinear nonconvex problem to solve, and it fails to guarantee a solution with minimized error.

*C. A probabilistic view: better than MLE*

Next, we compare the traditional SE method and the ECP-based approach from a probabilistic view, assuming the measurement errors satisfy independent Gaussian distribution $n_i \sim \mathcal{N}(0, \sigma_i^2)$. If we converge to the global optima of the defined non-convex optimization in first place, solution of the traditional WLS method is a maximum likelihood estimate (MLE), with its Lagrangian function equal to a log-likelihood:

$$L_{traditional} = \sum_{meter, i} w_i (z_i - f_i(x))^2 \quad (21)$$

Mathematically, MLE finds the model that is most likely to generate the data. This intrinsic property of MLE makes the traditional WLS method (a classical MLE) purely measurement-dependent: it will always make estimations purely based on measurements.

In contrast, from the Lagrangian function of the ECF-based method (still using example in Figure 5):

$$L_{ECF} = \sum_{bus\ i} w_i \left\| z_i' - Y_i x \right\|_2^2 + \sum_{s \in \{R, I\}} (\lambda_2^s Y_2^s x) \quad (22)$$

This ECF-based formulation maximizes a log likelihood plus a topology-based term. This additional term serves to include some accurate prior knowledge into the estimate, which is conceptually similar to a Bayesian treatment.

In general, in the non-Gaussian case, the topology-based terms in the objective in (22) may be viewed as a regularized M-estimator [2] that promotes physically relevant solutions (estimates). Thus, compared with the traditional one, the ECF-based method turns out to consider more physics and avoid extreme conclusions like those in (non-regularized) MLE.

Importantly, this viewpoint suggests a potential advantage for large-scale systems and distribution networks. As system grows larger, the increased incorporation of accurate real-world system physics will make more contributions to resulting in physically meaningful solutions.

IV. EXPERIMENTS

In this section we conduct experiments to compare the WLS-based SE method on MATPOWER[14] with the ECF-based method coded in python3. We create RTU measurements on each injection bus by adding Gaussian noise with standard deviation 0.001 to true |V| injections and line flows obtained by power flow simulation. To guarantee observability in the traditional approach, we modify the MATPOWER code to accommodate pseudo-measurements of zero P and Q injections for all zero injection (ZI) buses and weigh these measurements heavily. To make it run faster, we execute the MATPOWER code on Red Hat Enterprise operating system with 32-core Intel(R) Xeon(R) CPU @5-2620 v4 @2.10GHz.

We consider 3 standard metrics in (23)-(25) to evaluate the methods. Optimization residual $res$ measures the distance between observed meters and the estimation. This metric is exactly the objective function value which tells how well the algorithm converges. The second metric $MSE_z$ is mean weighted square error of $z$ which measures the distance between estimated meters and their true values. The third one $MSE_x$ is mean square error of state variable $x$, telling how close the solution is to the true state of grid. For evaluation of the state estimation methods, the error metrics in (24)-(25) are even more important than residual metric in (23), since a smaller deviation from truth is what we really care about.

$$res = \frac{\sum_i w_i (z_i - z_{est,i})^2}{n_{meter}} \quad (23)$$

$$MSE_z = \frac{\sum_i w_i (z_{est,i} - z_{true,i})^2}{n_{meter}} \quad (24)$$

$$MSE_x = \frac{\left\| x_{est} - x_{true} \right\|_2^2}{n_{bus}} \quad (25)$$

Results are shown in Table 3. The divergences in WLS-based method reflect that traditional iterative method relies highly on good initial guess. Clearly, the different starting points leads to different solution trajectories that vary in number of iterations and convergence. Generally, from Table 3, starting from initial conditions in the input file is better than flat start with 1 p.u. for all buses, but neither initialization assures convergence for all of our examples. Especially when the cases become larger, the traditional algorithm, without proper initialization, is very susceptible to divergence.

TABLE 3: METHOD COMPARISON RESULT

| Case Name | MATPOWER WLS-based SE | | | | | | | | ECF-base SE | | |
|---|---|---|---|---|---|---|---|---|---|---|---|
| | Initial guess from case data | | | | Flat initial guess | | | | | | |
| | $res$ | $MSE_z$ | $MSE_x$ | #of iter | $res$ | $MSE_z$ | $MSE_x$ | #of iter | $res$ | $MSE_z$ | $MSE_x$ |
| case14 | 199.8 | 198.6 | 0.00578 | 4 | 199.8 | 198.6 | 0.00578 | 5 | 0.82287 | 1.3489 | 0.00062 |
| case118 | 177.7 | 176.5 | 0.00350 | 4 | 177.7 | 176.5 | 0.00350 | 6 | 2.9901 | 3.4442 | 0.00348 |
| case2383wp | Diverge! Unobservability encountered. | | | | 210.6 | 209.8 | 0.01317 | 6 | 1.3872 | 1.9623 | 0.00139 |
| case3375wp | 438.35 | 438.36 | 0.04929 | 5 | Diverge! Unobservability encountered. | | | | 2.0155 | 2.5947 | 0.00152 |
| case6468rte | 3042.8 | 3041.7 | 0.04304 | 7 | Diverge! Unobservability encountered. | | | | 162.88 | 163.47 | 0.00793 |
| case9241pegase | 5886.45 | 5886.13 | 0.06532 | 11 | Diverge! Unobservability encountered. | | | | 312.78 | 313.35 | 0.01248 |
| ACTIVSg25k | No convergence within 6 hours. Unobservability encountered. | | | | | | | | 17.687 | 18.281 | 0.00371 |

Our ECF-based method, in contrast, doesn't need any initial guess or iterative updates and obtains a guaranteed global optimal solution in 1 step, resolving the convergence issues successfully. Particularly for larger cases, the guaranteed convergence and reduced runtime is extremely beneficial. Also, by comparison of residual metric $res$, we can see that our ECF-based method better minimizes the objective function.

Furthermore, comparison of error metrics $MSE_z$ and $MSE_x$ shows that our method reaches more accurate estimates. Specifically, the lower $MSE_x$ validates that our physics-based estimator produces a solution closer to the true operating point, enabling more reliable control actions to be made.

V. CONCLUSION

This paper presents a comprehensive comparison between traditional State Estimation and an ECF-based SE approach. The results show the ECF-based method addresses the root causes of SE limitations, and provides the following properties:

- **Linear to solve**. The SE problem reduces to a Quadratic Programming (QP) with closed-form solution, without any concern about initial guess and convergence issue.
- **More physically meaningful estimation** by equivalently imposing additional topology-based constraints on the WLS problem. These constraints provide a structure-based regularization effect by shrinking the feasible space to promote convergence to an optimal solution that not only tends to minimize the weighted squared error but also to satisfy network balance.
- **Incorporation of prior knowledge into the estimate**, better than the MLE that is dependent solely on measurements and likely to give extreme nonphysical solutions.

These advantages make the ECF-based approach particularly useful for large-scale systems and other SE-related topics such as bad-data detection.

VI. ACKNOWLEDGMENT

This work was supported in part by the Defense Advanced Research Projects Agency (DARPA) under award FA8750-17-1-0059 for the RADICS program, and the National Science Foundation (NSF) under contract ECCS-1800812. We would also like to thank Lydia Hays for discussions regarding the current source-based PMU model that was used in this paper.